\documentstyle[aps,prl]{revtex}

\newcommand{\BE}{\begin{equation}}
\newcommand{\EE}{\end{equation}}
\newcommand{\BA}{\begin{eqnarray}}
\newcommand{\EA}{\end{eqnarray}}

\begin{document}
\draft

\twocolumn[\hsize\textwidth\columnwidth\hsize\csname@twocolumnfalse\endcsname
\title{Phase Transition and Acoustic Localization in Arrays of Air-Bubbles in Water}
\author{Zhen Ye\footnote{Email: zhen@joule.phy.ncu.edu.tw}
and Haoran Hsu}
\address{Department of Physics and
Center for Complex Systems,
National Central University, Chungli, Taiwan, ROC}
\date{Feb. 5, 1999}

\maketitle

\begin{abstract}

Wave localization is a ubiquitous phenomenon. It refers to
situations that transmitted waves in scattering media are trapped
in space and remain confined in the vicinity of the initial site
until dissipated. Here we report a phase transition from
acoustically extended to localized states in arrays of identical
air-filled bubbles in water. It is shown that the acoustic
localization in such media is coincident with the complete band
gap of a lattice arrangement of the air-bubbles. When the
localization or the band gap occurs, a peculiar collective
behavior of the bubbles appears.

\end{abstract}

]


Under appropriate conditions, multiple scattering of waves
leads to the interesting phenomenon of wave localization,
which has been
and continues to be a subject of substantial
research(e.~g. Refs.~\cite{EPL,RMP,Nature1,Nature}).
It has been shown\cite{EPL} that localization
can be achieved for acoustic waves propagation in liquids with even a very
small fraction of air-filled spherical bubbles. It is
shown that the localization appears within a range of frequency slightly
above the natural resonance of the individual air bubbles. Outside this
region, wave propagation remains extended.

Multiple scattering
and Bragg refraction of acoustic
waves in periodic structures can also lead
to what has been called a `complete band gap', which is another
fascinating research subject\cite{K1}. In the gaps, no wave can
propagate in any direction, a useful feature for noise isolation.
In this paper we mainly consider acoustic wave localization in
random arrays of air-bubbles in water and its possible connections to
the complete band gaps of a crystal array of these air-bubbles.
We show a coincidence between the localization and
the complete band gap in such bubbly liquids.
A novel phase diagram is used to characterize the phase transition from the
extended to localized states. When the phase transition occurs, an
interesting coherent behavior appears.

Consider sound emission from a unit acoustic
source located at the center of a bubble ensemble.
The acoustic source transmits a
monochromatic wave of angular frequency $\omega$.
Total number $N$ spherical bubbles of the same radius $a$ are
either randomly placed in water or placed regularly to
form a simple cubic lattice with lattice constant $d$.
In either cases, the bubble volume fraction, the space occupied by bubbles
per unit volume, is the same and taken as $\beta (= a^3/d^3)$.
The coordinates of the bubbles are
denoted by $\vec{r}_i$ with $i = 1, 2, \dots, N$.
The wave transmitted from
the source propagates through the bubbly structure,
where multiple scattering incurs, and then it reaches a receiver located at
some distance from the bubble ensemble. The multiple scattering
is described by a set of self-consistent equations\cite{Foldy} and
can be solved rigorously\cite{PRL}.

It is found that the localization of acoustic waves is reached when the
bubble volume fraction $\beta$ is greater than a threshold value of
10$^{-5}$ and the localization behavior is insensitive
to the bubble size when the bubble radius is larger than 20
$\mu$m\cite{PRL}.
Note that for smaller bubbles, the thermal and viscosity effects are
significant enough that the wave localization seems less evident.
In the simulation,
we take $\beta = 10^{-3}$, and the total number of bubbles is varied
from 100 to 2500, large enough to eliminate possible effects on localized
states due to the finite sample size. The radius of the bubbles ranges from
20 $\mu$ to 2 cm.

The scattered wave from each bubble is a linear
response to the direct incident wave $p_0$ from the source and also
{\it all} the scattered waves $p_s$ from other scatterers, and is
written as\cite{PRL}
\BE
p_s(\vec{r}, \vec{r}_i) = f_i\left(p_0(\vec{r}_i) +
\sum_{j = 1, j\neq i}^{N} p_s(\vec{r}_i, \vec{r}_j)\right)
G_0(\vec{r}-\vec{r}_i), \label{eq:1}
\EE
where $f_i$ is the scattering function of a single bubble, and
$G_0(\vec{r}) = \exp(ikr)/r$ is the usual 3D Green's function with
$k$ being the usual wavenumber. The total wave $p(\vec{r})$ at
any space point is the summation
of the direct wave from the transmitting source and all the scattered waves.
The scattering function $f_i$ can be readily computed and the solution
can be written in the form of a modal series, representing various
vibrational modes of each spherical bubble. In the frequency range
considered, the pulsating mode dominates the scattering. To solve for
$p_s(\vec{r}_j, \vec{r}_i)$ and subsequently $p(\vec{r})$, we
set $\vec{r}$ in Eq.~(\ref{eq:1}) to one of the
scatterers other than the $i$-th bubble. Then Eq.~(\ref{eq:1})
becomes a set of closed
self-consistent equations which can be solved exactly by matrix inversion.
We define $I = <|p|^2>$ to represent the squared modulus of the total
waves, corresponding to the total energy.

A set of numerical experiments has been carried out for various bubble
sizes, numbers, and concentrations. Fig.~1 presents one of the typical
results of the
total wave transmission as a function of frequency in terms of $ka$ for
one random configuration of the bubble cloud.
As shown, the transmission is greatly reduced
roughly between 0.016 and 0.08 in $ka$, indicating the localization regime.
Note that a few peaks appear below the localization regime.
The peak at the lowest frequency is attributed to the effect of finite sample
size, and will gradually disappear as the sample size is enlarged. The
peak closest to the localization regime is due to the
resonance of a single bubble embedded in the ensemble.

Upon incidence, each air-bubble acts effectively as a secondary
pulsating source. The scattered wave from the $i$-th bubble
($i = 1, 2, 3,..., N$) is regarded as the radiated wave, and
can be rewritten as $A_iG_0(\vec{r}-\vec{r}_i)$. The complex coefficient
$A_i$ is computed incorporating {\it all} multiple scattering effects.
Express $A_i$ as $|A_i|\exp(j\theta_i)$ with $j = \sqrt{-1}$ here;
the modulus $A_i$
represents the strength, whereas $\theta_i$ the phase of the secondary
source. We assign a two dimensional unit vector $\vec{u}_i$, hereafter termed
phase vector, to each phase $\theta_i$, and these vectors are
represented on a phase diagram parallel to the $x-y$ plane. That is, the
starting point of each phase vector is positioned at the center of
individual scatterers with an angle with respect to the positive
$x$-axis equal to the phase, $\vec{u}_i = \cos\theta_i\hat{x} +
\sin\theta_i\hat{y}$.
Letting the phase of the initiative emitting source be zero,
i.~e. the phase vector of the source is pointing to the positive
$x$-direction,
numerical experiments are carried out to study the behavior of the phases
of the bubbles and the spatial distribution of the acoustic wave modulus.
The magnitude of the summation of all phase vectors may represent an
order parameter.

Figure~2 shows the phase diagrams of the phase vectors (left
column) and the averaged
wave distribution as a function of distance from the source (right column).
The wave distribution is plotted in arbitrary scale and the
uninteresting geometric spreading is removed.
Three particular frequencies are chosen
below, within and above the localization regime,
according to the results of Ref.~\cite{PRL}: $ka = 0.01254, 0.01639$, and
$0.1$. The dots refer to the three dimensional positions of the air
bubbles at $\vec{r}_i = (x_i, y_i, z_i)$, with $i = 1, 2, \dots, N$.
The arrows refer to the phase vectors which are
located at the bubble sites and are parallel to the $x-y$ plane.
The distance dependence of the total wave is scaled by the
sample size $R (\sim N^{1/3}d)$. For the
purpose of demonstrating the physical picture in its most explicit way,
we have only shown here the phase vectors for 136 bubbles.

We observe that for frequencies below the localization regime, roughly below
$ka = 0.016$,
there is no obvious ordering for the directions of the phase vectors, nor
for the wave distribution. The phase vectors point to various
directions; the order parameter is nearly
zero. In this case, no wave localization appears, corresponding to
the extended state in Fig.~1.
These features are shown by
the example of $ka = 0.01254$, roughly at the resonance of the single
bubble. Due to the finite sample size, an oscillatory wave
distribution is observed inside the sample, in agreement with a
computation from the mean field theory that regards the bubbly water as
a uniform medium.

As the frequency increases, moving into the
localization regime, the wave localization and
an ordering of the phase vectors
becomes evident. The case with $ka = 0.01639$
clearly shows that the wave is localized near the source.
In the meantime, all bubbles oscillate completely in phase, but
exactly out of phase with the transmitting source; all phase vectors point
to the negative $x$-axis, in parallel to the $x-y$ plane.
Such a collective behavior allows for efficient cancellation of
propagating waves. The appearance of the
collective behavior for localized waves implies existence of a kind of
Goldstone bosons incurred in the phase transition.
The total wave (energy)
decays nearly exponentially along the
distance of propagation inside the bubble ensemble,
setting the localization length\cite{PRL}. Outside the sample, the
transmission remains constant with distance, as expected.
The wave localization and the order in the phase vectors are
independent of the outer boundary and {\it always} appear for sufficiently
large $\beta$ and $N$.

When the frequency increases further,
moving out of the wave localization region, the
in-phase order disappears. Meanwhile, the wave becomes non-localized again.
This is illustrated by the case of $ka = 0.1$. In this case, the phase
vectors again point to various directions and the wave distributes roughly
uniformly in space. A further study shows that at this frequency,
the multiple scattering is negligible. Each bubble
mainly experiences the direct incident wave from the source. Thus the phase
of each bubble approximates the phase of the incident wave
plus a constant phase from the response function of the bubble. As a
result, the bubbles at equal distance from the source have the same
phase vectors.

In exploring the relation between the above localization behavior
with the band gap of regular bubble arrays, we have computed
the band structure of a
cubic array of air bubbles following Ref.~\cite{K1}.
The result is shown in Fig.~3. We observe that there is a complete band gap
between $ka = 0.017$ to 0.09, which coincides with the localization
range shown by Fig.~1. The phase vectors and the energy distribution
in this cubic array of bubbles are also plotted in Fig.~4 for the three
frequencies chosen above. The similar features appear: outside the gap
the phase vectors point to various directions, whereas inside the
gap all phase vectors point to the same direction in the opposite of
the phase vector of the source. As expected, the energy is extended
for frequencies outside the gap and localized inside the gap.
The similarity between the cases of the random and ordered arrays
implies the same mechanism for both the localization and
the complete band gap in such systems, and the phase
transition between extended and forbidden bands is similar to that
between non-localization and localization.

An intuitive understanding about the acoustic localization
in the bubbly structures may be helped from the following consensus.
Imagine that two persons hold a thread. If one person pushes, while the other
pulls, the two persons would act completely out of phase. No energy
can be transferred from one to the other. In the bubbly liquid case, the
state of localized waves shows such a completely out-of-phase behavior, which
effectively prevents waves from propagating.

We have demonstrated a new phase transition for acoustic
propagation in arrays of air-bubbles
in water. The results show that
as the phase transition occurs, not only the acoustic waves are
confined in
the neighborhood of the transmitting source, but an amazing
collective
behavior of the air bubbles appears. The result suggests
a coincidence between wave localization and complete wave band
gaps.

\noindent {\bf Acknowledgments.}
Discussion and help from Alberto Alvarez and Emile Hoskinson are
appreciated.
The work received support from National Science Council of ROC.


\section*{Figure Captions}

\begin{description}

\item[Figure 1] Averaged transmission as a function of $ka$.

\item[Figure 2] Left column: The phase diagram for the two dimensional
phase vectors defined in the text for one random distribution of
bubbles. Right column: The acoustic energy distribution or
transmission, averaged for all directions, as a function of
distance away from the source; the distance is scaled by the
sample size $R$. ({\bf Original in Color})

\item[Figure 3] The acoustic band structure for a simple cubic
array of air-bubbles. The complete band gap lies between the two
dashed horizontal lines. The plot is rendered in terms of the
non-dimensional $ka$ as a function of Bloch vectors.

\item[Figure 4] The phase diagram for the two dimensional phase
vectors and the energy distribution, averaged for all directions,
for the cubic array of the bubbles which has the same bubble
volume fraction as the random case. The lattice constant $d =
1/(\beta)^{1/3}a$. ({\bf Original in Color})

\end{description}

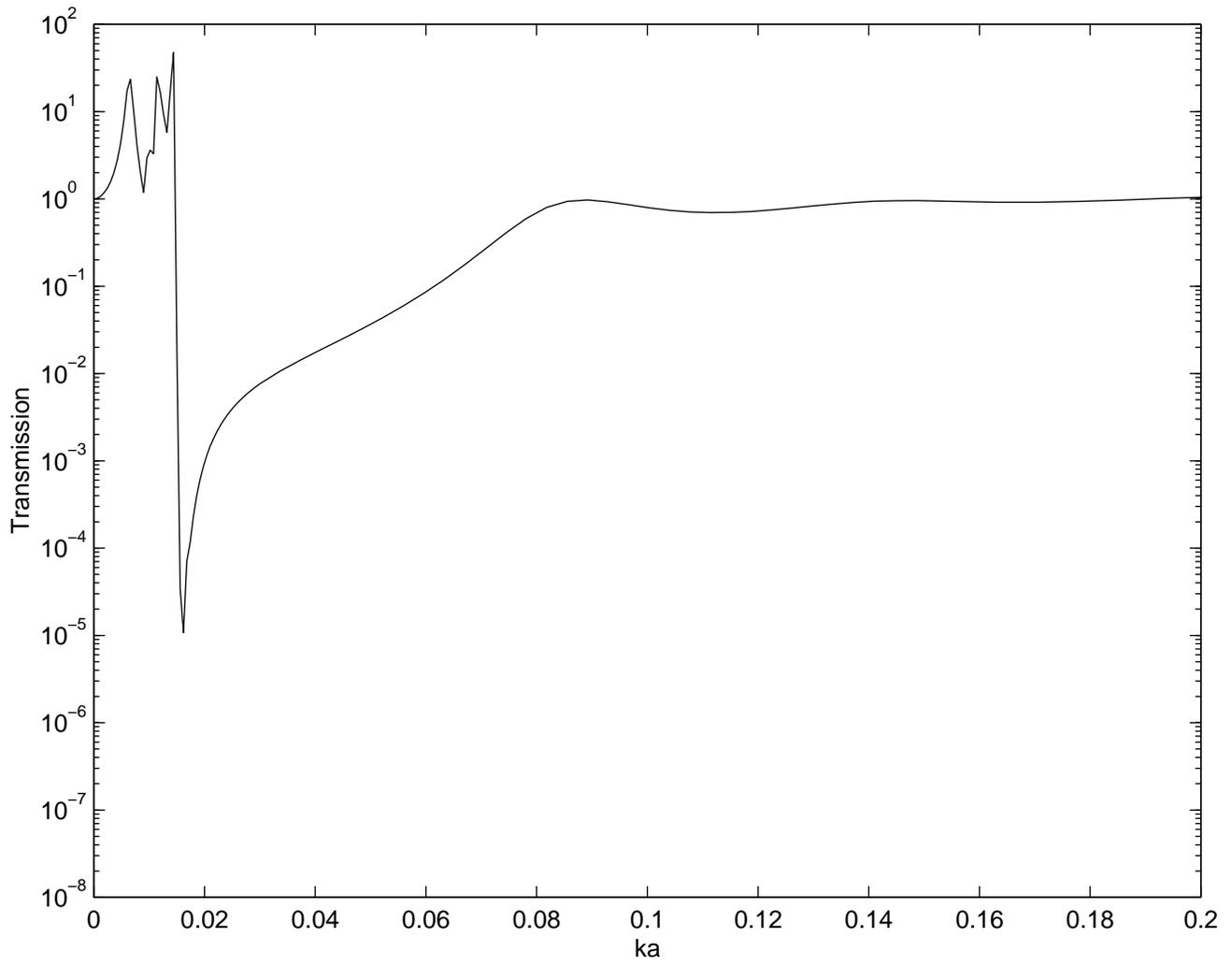
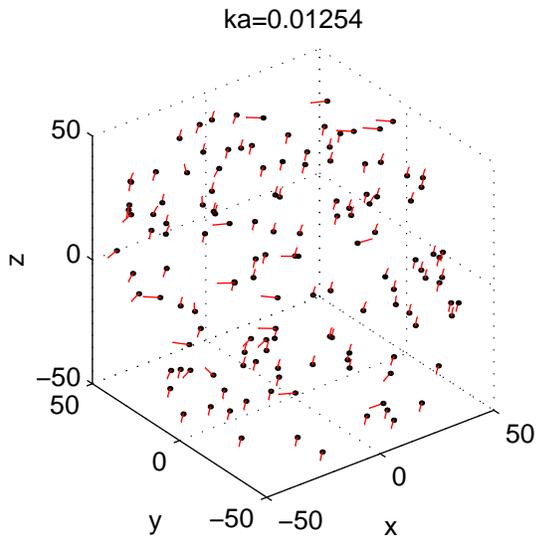
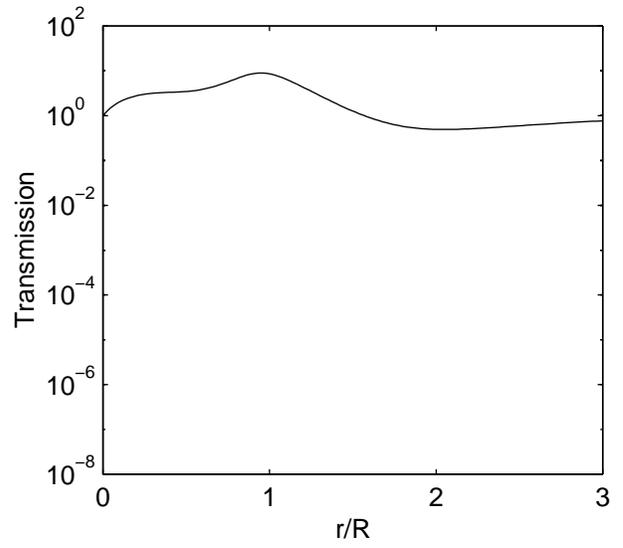
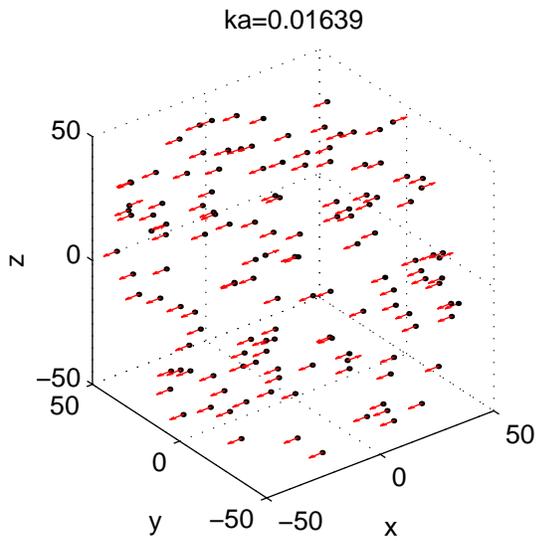
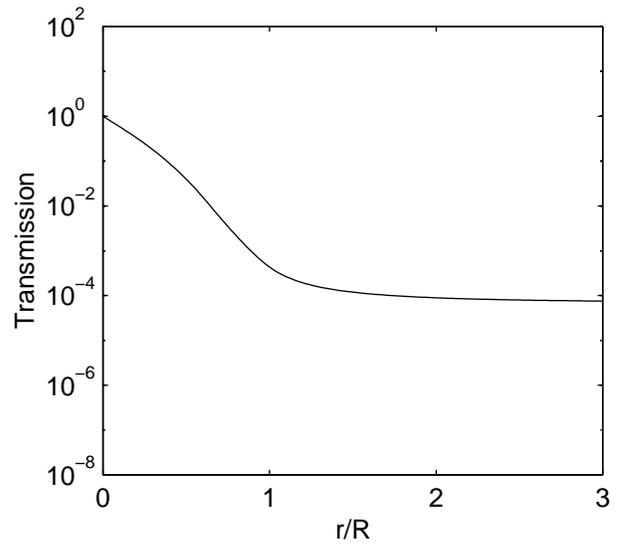
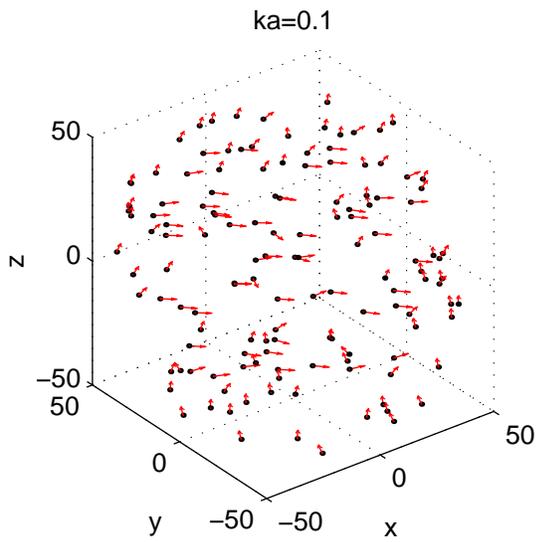
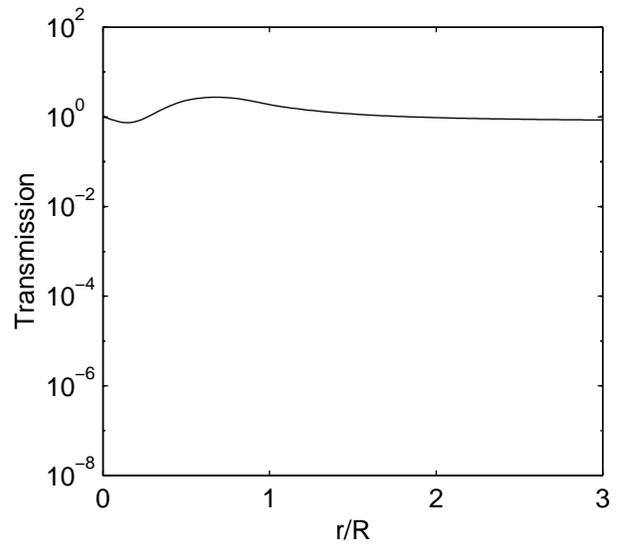
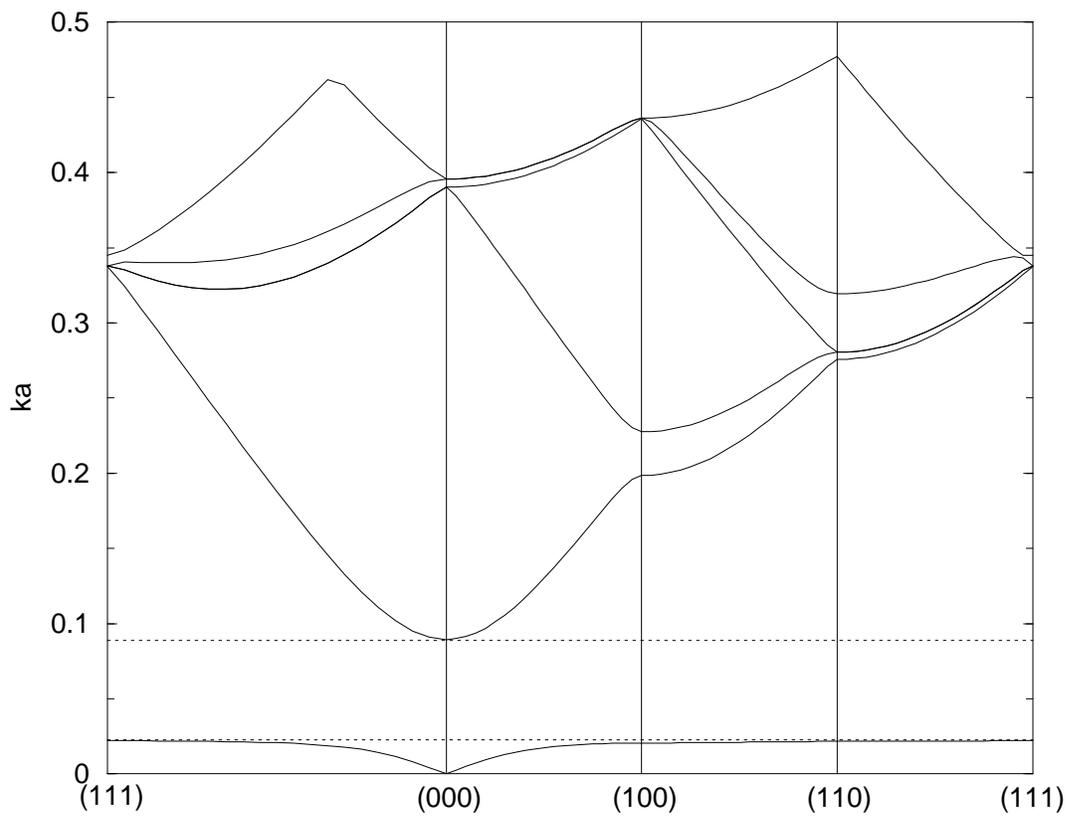
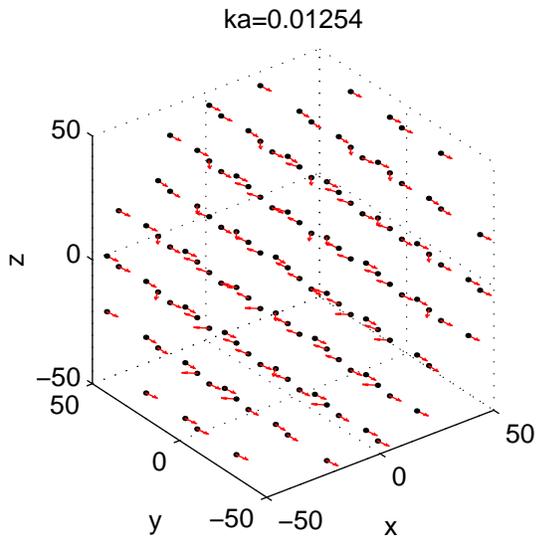
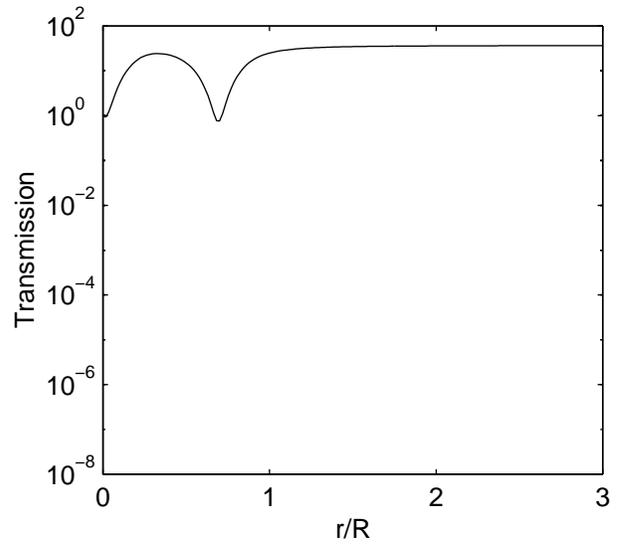
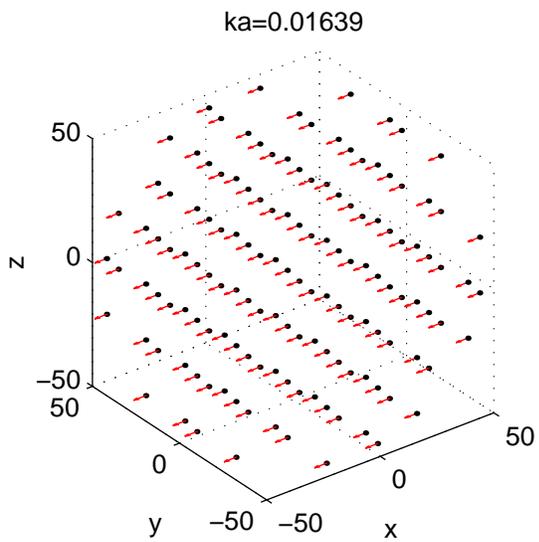
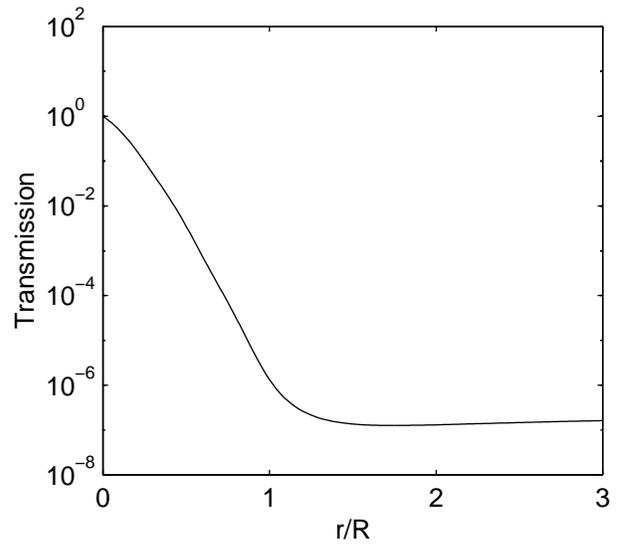
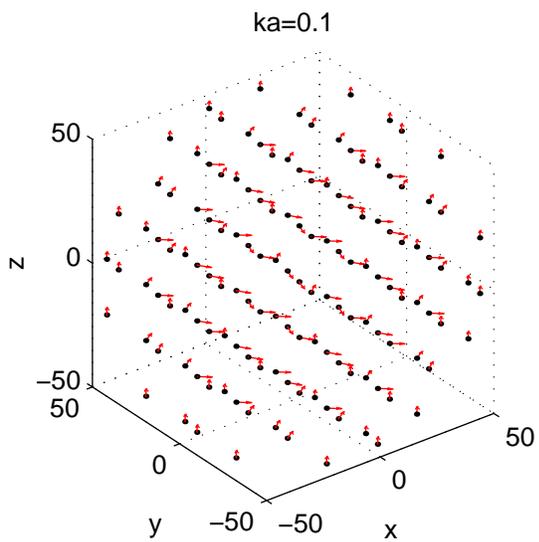
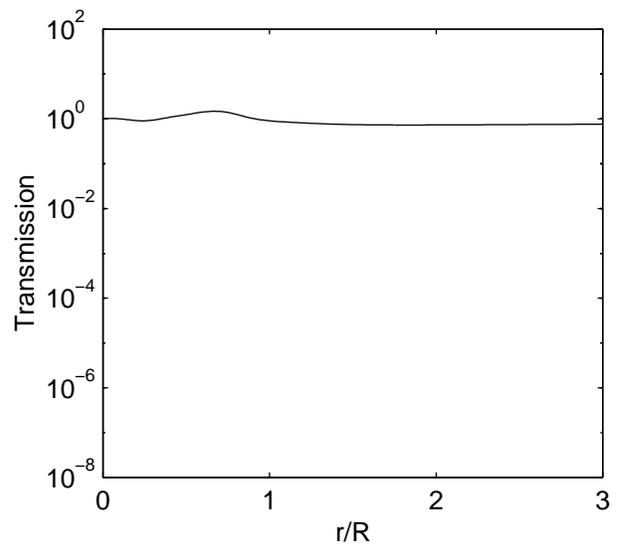
\end{document}